\documentclass[11pt,a4paper]{JHEP}
\usepackage{amsmath}
\usepackage{amsfonts}
\usepackage{amssymb}

\def\v{{\rm v}}

\def\d{\partial}

\newcommand\0{\nonumber}
\newcommand\ee{\end{eqnarray}}	 	
\newcommand\be{\begin{eqnarray}}
\newcommand\ba{\begin{array}}			
\newcommand\ea{\end{array}}

\preprint{SISSA/136/99/EP\\hep-th/9912214}

\title{Normal Bundles, Pfaffians and Anomalies}

\author{ L.\ Bonora\\
International School for Advanced Studies (SISSA/ISAS)\\
Via Beirut 2--4, 34014 Trieste, Italy, and INFN, Sezione di Trieste\\
E-mail:   \email{bonora@sissa.it}, }
\author{M.Rinaldi\\Facolt\`a di Farmacia, Universit\`a del Piemonte
Orientale\\
Viale F. Ferrucci 33, 28100 Novara,Italy\\
E-mail: \email{rinaldi@pharm.unipmn.it}}

\abstract{We deal with the problem of diffeomorphism anomaly in theories with
branes. In particular we thoroughly analyze the problem of the residual
chiral anomaly of a five--brane immersed in M--theory, paying attention to 
its global formulation in the five--brane world--volume. We conclude that 
the anomaly can be canceled by a {\it local} counterterm in the five--brane 
world--volume.}


\begin{document}

\section{Introduction}

This article deals with the problem of diffeomorphism anomalies
in theories with branes. With respect to the traditional anomaly analysis in
field or superstring theories without branes, the presence of branes
introduces new questions.
Macroscopic branes in superstring or M--theory are represented, from a
geometrical point of view, by submanifold embedded in the 10 or
11 dimensional ambient manifold. On the brane world--volumes there live fields
that represent the dynamical degrees of freedom of the brane.
Simultaneously the brane interacts with the ambient theory.
Let us suppose that the latter is anomaly free and that the fermionic
degrees of freedom on the brane are chiral. Then the overall theory
of the brane embedded in the ambient theory might contain chiral
anomalies which break the invariance under those diffeomorphisms that map the
brane world--volume to itself. Anomaly contributions may be of three types:
there may be
anomalies of the brane theory in isolation, anomalies due to the embedding,
i.e. anomalies due to pulled--back metrics or connections from the ambient
manifold, and finally inflow anomalies, i.e. anomalies due to the
interaction of the brane with the ambient theory. All these types
contain contributions from the tangent bundle of the world--volume $W$
of the embedded brane, but the last two types
of anomalies contain also normal bundle contributions. They come in fact from
characteristic classes of the
ambient tangent bundle $TM$, where $M$ is the ambient space--time;
the well--known decomposition $\left. TM\right\vert_W= TW\oplus N$ holds,
$N$ being the bundle normal to the brane and the characteristic classes of
$TM$
will split accordingly. The purely $TW$ part of the anomaly can be
thought of as due to diffeomorphisms that map $W\to W$. On the other hand, the
$N$ part of the anomaly
can be thought of as due to diffeomorphisms that leave $W$ pointwise fixed and
can be interpreted as gauge transformations of $N$. These two cases have an
easy
geometrical interpretation. They are essentially the two cases considered
so far
in the literature. However they do not represent the most general situation.
Apart from this, in all the examples considered so far, only
local expressions of the relevant anomalies have been used. Reality is more
complicated than this. We need to refine the analysis in at least two
directions.

To start with there are other diffeomorphisms of $M$, which map $W$ to $W$ but
are not comprised in the two subgroups mentioned above. We remark that if
we apply
a diffeomorphism that maps $W$ to $W$, in general the normal bundle
is deformed and not mapped to itself. We will argue later on that one must
consider only diffeomorphisms which map $N$ (globally) to itself. Even with
this
simplification the relevant anomalies cannot be thought of
and treated as usual gauge anomalies. One has to rely on a more general
formalism which involves both diffeomorphisms and gauge transformations on
the same footing. This can be done by considering general automorphisms of the
relevant principal bundle (rather than vertical automorphisms, i.e.
ordinary gauge transformations, alone).

In addition we will need formulas for anomalies expressed as basic
forms in the space--time manifold $M$. The only way to achieve this
is by introducing a background connection, i.e. a spectator connection
which is not transformed under the relevant transformations. Formulas for
anomalies with a background connection can be found in the literature,
\cite{MSZ,BCRS1}. Introducing a background connection is allowed
for the following reason. Let us consider a principal fiber bundle $P(M,G)$ and
a given connection $A$. Any automorphism $\psi\in Aut P$ maps $P$ to $P$ but
`rotates' it, i.e. $P \to\psi^{*}P$, and maps any connection $A$ to $\psi^*A$.
However $A$ is a
connection both in the original bundle and in the transformed one. Therefore
it is
consistent to keep one connection $A_0$ fixed, while rotating the bundle and
all the other connections. The background connection $A_0$ allows us to write
down local expressions for anomalies. This is of course not in contradiction
with the usual local expressions of the same anomalies, which can be recovered
by simply setting (locally) $A_0=0$. As we shall see, the framework
considered in the paper needs in fact further qualifications with respect to the
ordinary gauge theories setting just described.

As far as anomalies are concerned, however, one does not expect any
significant complication
from this more general treatment whenever anomaly cancellation takes place
at the level of characteristic classes (apart, of course, for the necessity to
specify an appropriate geometrical setup). For in this case anomalies
are canceled at the source, so to speak, and there is nothing left on
which an anomalous behaviour can build up. However the situation is
different when a residual anomaly is canceled by means of a mechanism \'a la
Green--Schwarz. In the latter case there is a physical input (the existence
of a suitable local field) that does not follow simply from the automatisms
of the descent equations.

In this regard there is a gap in the analysis carried out so far in the
literature on the subject. In \cite{BR} we have shown how to implement the
anomaly cancellation mechanism with background connection in the original
Green--Schwarz case. In this paper we wish to extend the analysis to theories
with branes. Actually we will concentrate on the case of the M--5--brane
anomaly,
which alone contains all the above complications: it is a normal bundle anomaly
which can only be canceled via a
Green--Schwarz mechanism, \cite{witten,alwis,henningson,BCR,BB} and
\cite{FHMM}
(analogous problems arise in other cases, \cite{mourad,braxmourad,CY},
which will
not be discussed here but can be treated along the same lines). 

Let us summarize the M--5--brane problem. The geometric
setting for this problem, \cite{witten},
is specified by the 11d manifold $X$ of M--theory and by the 6d manifold
$W$, representing the world--volume of the 5--brane embedded in it.
In addition we have
the well--known decomposition $\left.TX\right\vert_W= TX\oplus N$, $N$
being the
bundle normal to the brane world--volume, whose structure group is $SO(5)$.
In isolation, both theories on $X$ and on $W$ are anomaly--free. But,
due to the embedding of $W$ in $X$ and to the physical coupling of the
5--brane (see eq.(\ref{df4}) below), one gets both induced and inflow anomalies
on $W$. These contributions to the anomaly do not cancel completely:
the residual anomaly is generated via the descent equations by
the 8--form $\frac {1}{24}p_2(N)$, where $p_2(N)$ represents
the second Pontryagin class of the normal bundle.
Now, the brane is magnetically coupled to M--theory via
\be
dF_4= \delta_W\label{df4}
\ee
where $\delta_W$ is a representative of the Poincar\'e dual of $W$, and $F_4$
is the 4--form field strength of 11d supergravity. We showed
in \cite{BCR} that, due to (\ref{df4}), $W=\d Y$, i.e. $W$ is the boundary
of some
7--manifold $Y$. This implies that the normal bundle $N$ is a direct sum,
$L\oplus N'$, of a trivial line bundle $L$ and a vector bundle $N'$ whose
structure group is reduced to $SO(4)$. It follows that
the second Pontryagin class of $N$ becomes:
\be
p_2(N)= p_2(N')= e(N')^2\equiv e^2\label{e2}
\ee
where $e$ is the Euler class of the normal bundle. Therefore the anomaly
we have to do with is generated via the descent equations by $e$. This
is, roughly speaking, a summary of the M--5--brane anomaly problem.
We will clarify, in the course of the paper, various aspects of this problem.
But we would like to emphasize from the start the aspect of locality.
Throughout the paper the form $\delta_W$ in (\ref{df4}) is taken to be the 
Dirac--delta 5--form. This allows one to work within a {\it local} field theory framework (with a topological defect). Since the residual anomaly has a 
local expression on the brane world--volume, one has to require that it 
be canceled by a local counterterm. From this point of view, however, 
the existing literature does not offer a satisfactory solution. This is the
main problem we want to cope with.
 
In the present paper we intend to fill in the gaps described above.
We propose a {\it local} counterterm to cancel the residual
M--5--brane anomaly. This mechanism is tailored to take into account
eq.(\ref{df4}). It is worth insisting that this equation, which expresses
the magnetic coupling of the 5--brane to M--theory,
requires the normal bundle splitting $N=L\oplus N'$: this must be reflected
both in the form of the anomaly and of the counterterm. We will see that this
splitting is essential for the anomaly cancellation. 

While meeting the above requirements, we make 
a point of using basic expressions both for the anomaly and the counterterm and
properly take into account the problems connected with diffeomorphisms 
in the presence of a normal bundle.

The paper is organized as follows. The next section is preliminary to the
anomaly analysis. Since a crucial role in our problem is played by the
reduction from $SO(5)$ bundles and connections to $SO(4)$ ones, we devote
a few pages to deriving a workable formalism to deal with reducible
connections. In particular we find for them an explicit formula,
(\ref{redconn}),
in terms of a unimodular section $v$ which defines the corresponding reduction.
Next we show that for reducible connections the form which represents
the second Pontryagin class factorizes into the square of a Pfaffian. In
section
3 we set out to calculate the expression of the anomaly corresponding to such
a factorized class, and succeed in finding an expression for it which is
globally defined on the world--volume of the 5--brane. Finally in section 4
we introduce the local and globally defined counterterm which cancels such
an anomaly. We discuss the meaning of the counterterm and of the field
variables $v$ introduced to parametrize the various reductions.

\section{Reducible connections, Pfaffians and Anomalies}

Our purpose in this section is to find a basic expression of the residual
anomaly of the M--5--brane, taking into account the two complications
mentioned in the introduction.

\subsection{The geometric setting}

The first task is to specify what subgroup of the diffeomorphisms of $X$ is
to be considered. It would seem natural to consider diffeomorphisms of $X$
which
map $W$ to itself. However we notice that, while $TW$ is mapped to itself by
any diffeomorphism of this kind, the same is not true for the normal bundle:
if $\psi$ is any such diffeomorphism, $N$ and $\psi^*N$ are in general
different subbundles of $TX$. Now, while a generic connection
$A$ in $N$ is mapped to a connection $\psi^*A$ in $\psi^*N$, this is not
true anymore for a background connection $A_0$. Recalling the above remarks on
background connections we point out that, if $A_0$ is a fixed connection
of $N$, it cannot in general be a connection in $\psi^*N$. If we want a basic
expression of the anomaly, we have to come to terms with this fact. Therefore
we restrict the subgroup of allowed diffeomorphisms to those that leave $N$
globally invariant and denote it as $Diff(X,N)$. This implies that $N$ has
some
physical significance. In fact the M--5--brane spectrum contains five scalar
fields, which span the normal directions to the 5--brane. A symmetry
transformation permitted by physics can only transform each one of five fields
into a combination of them, but will not be allowed to switch on new
directions.
This is exactly what the global invariance of $N$ means.

The subgroup $Diff(X,N)$ can also be seen more
usefully as $Aut P$, the group of automorphisms of the principal fiber bundle
$P$ with structure group $SO(5)$, associated to the normal bundle $N$. In this
paper we will consider only the infinitesimal version, its Lie algebra
$aut P$. The anomaly of the M--5--brane is the anomaly with respect to
the transformations $Z$, which are the vector fields
in $aut P$.
The next thing to consider is the splitting $N= L\oplus N'$. This is an
inevitable consequence of (\ref{df4}), see \cite{BCR}, and induces the
reduction of the
structure group from $SO(5)$ to $SO(4)$. There is a manifold of such splittings
and one can take two attitudes: either one assumes there is a privileged one
and then further limit $Diff(X,N)$ to those diffeomorphisms that preserve
such a splitting, or else one takes all of them into account. We will resume
later on this discussion. For the time being we treat the problem in complete
generality: first of all, we classify all possible reductions of $SO(5)$
to $SO(4)$; then we notice that a diffeomorphism of $Diff(X,N)$ maps a
reduced bundle into another reduced bundle, on the other hand the transform
of a
non--reducible connection in the first bundle is not a connection in the
transformed one; so it makes sense to consider in
$p_2(N)$ only reducible connections. Therefore we write down a general
formula for
reducible connections and show that $p_2(N)$ decomposes into the square of a
Pfaffian; finally we derive a basic formula for the anomaly.
In the next section, we show that it is possible to cancel it with a local
counterterm in $W$.

\subsection{The role of gamma matrices}

A relevant role in the following is played by gamma matrices and by the 
relation
between the fundamental representation of $\mathfrak{sp}(2)$ and the
 representation
${\bf 4}$ of $\mathfrak{so}(5)$. We devote the present subsection to these
pedagogical topics.
 
Gamma matrices for $\mathfrak{spin(5)}=\mathfrak{so}(5)$ are defined as follows:
take Euclidean gamma
matrices  in 4D, $\gamma_1,\ldots,\gamma_4$ and define
$\gamma_5=\gamma_1\gamma_2\gamma_3
\gamma_4$. Then $\gamma_a$ with $a=1,\ldots,5$ satisfy
$$\{\gamma_a,\gamma_b\}= 2\delta_{ab}$$
They form a $4\times 4$ matrix representation of the Clifford algebra of
$\mathbb R^5$.
We have $\gamma_a^2=1$ and
$$
{\rm Tr} (\gamma_a\gamma_b\gamma_c\gamma_d\gamma_e) = 4\epsilon_{abcde}
$$
with $\epsilon_{12345}=1$. The quadratic combinations
$$
\Sigma_{ab}= \frac{1}{4} [\gamma_a,\gamma_b]
$$
satisfy
\be
&&[\Sigma_{ab},\Sigma_{cd}]= -\delta_{ac}\Sigma_{bd}+\delta_{ad}\Sigma_{bc}
-\delta_{bd}\Sigma_{ac}+\delta_{bc}\Sigma_{ad}\nonumber\\
&&[\Sigma_{ab},\gamma_c]= \delta_{bc}\gamma_a- \delta_{ac}\gamma_b\nonumber
\ee
Therefore they are the generators of the representation ${\bf 4}$ of
$\mathfrak{so}(5)$.
The latter can be identified with the fundamental representation of
$\mathfrak{sp}(2)$ via the Lie-algebra isomorphism $\mathfrak{so}(5)\approx
\mathfrak{sp}(2)$.
Consider now the adjoint representation
$$\text{\rm Ad}\colon Spin(5)\to SO(5)$$
The associated Lie algebra isomorphism
\begin{equation}\text{\rm Ad}_*\colon \mathfrak {spin(5)}\to
\mathfrak{so}(5)\label{iso}\end{equation} is given on the basis elements
$\Sigma_{ij}=\frac
14[\gamma_i,\gamma_j]$ ($i<j$) by
$$\Sigma_{ij}\to   e_i\wedge e_j$$
Notice that $\mathfrak{so}(5)$ acts on vectors of $\mathbb R^5$, in the usual way
$$(e_i\wedge e_j)e_k=e_i\delta_{jk}-e_j\delta_{ik}$$ and so $e_i\wedge e_j$
can be identified with
the $5\times 5$ matrix
$(E_{ij})_{lm}=\delta_{il}\delta_{jm}-\delta_{im}\delta_{lj}$. Therefore
the corresponding representations can be implemented at the level of
$\mathfrak{spin}(5)$ on
$\Delta=\text{span}(\gamma_i)\subset \text{Mat}(4)$ as follows
$\Sigma_{ij}\cdot \gamma_k=[\Sigma_{ij},\gamma_k]$.    Therefore
for any vector $v^a$, $a=1,\ldots, 5$, it is convenient to define
$\v = v^a \gamma_a,$ i.e.
$v^a = \frac{1}{4}{\rm Tr} (\gamma^a \v)$.

\subsection{Reducible connections}

Our purpose in this subsection is to justify formula (\ref{redconn}) for 
reducible connections. On first reading one can jump directly to that 
formula and to the next subsection.

Consider the principal bundle $P\to W$ associated to the normal bundle
$N$  and
a reduction $  R\rightarrow W$ with structure group $SO(4)$ with
associated bundles  $N'$  and fiber $\mathbb R^4$.
Any reduction $j\colon R\rightarrow P$ corresponds
to a section  $\sigma\colon W\to E$  of the bundle $E$ associated to $P$
with fiber
$\frac{SO(5)}{SO(4)}$.  We will identify
$\frac{SO(5)}{SO(4)}=S^4$. To make this identification  precise  consider
the vector
$e=(0,0,0,0,1)\in
\mathbb R^5$ and let $j\colon SO(4)\to SO(5)$ be given by
$$j(h)=\pmatrix h&0\cr 0&1\endpmatrix$$
Then $SO(4)e=e$ and we map   $[g]\in \frac{SO(5)}{SO(4)}$  to $ge\in S^4$.
Once this
identification is made it is easy to identify $E$ with the unit sphere
subbundle $S(N)$ of the
normal bundle
$N$. To the section
$\sigma\colon W\to S(N)$ we can associate as usually a unimodular   map
$v\colon P\to
\mathbb R^5$, such that $v(pg)=g^{-1}v(p)$.
In fact
any point $p\in P$  (i.e. a basis $p=(X_1,\ldots,,X_5)$ of    $N_{\pi(p)}$)
defines a map
$\widehat p\colon \mathbb R^5\to N_{\pi(p)}M$ via $\widehat p(w)=\sum_{i=1}^5
X_iw_i$. Now define a function $v(p)\equiv \widehat p^{-1}\sigma(\pi(p))$.

Notice that $v$ determines a line bundle $L_v$ within the normal bundle
$N$ and
thus a splitting $N= L_v \oplus N_v'$, where $N_v'$ has structure group
$SO(4)$.

 Let now $A$ be a connection in $P$.
 By definition    $d_Av$ is a tensorial 1-form on $P$ with values in
$\mathbb R^5$. Consider the 1-form $\omega=v\wedge d_Av$ on $P$. 
Given  the canonical basis $e_i$
of $\mathbb R^5$ we can decompose $v$ as $v=\sum_{i=1}^5 v_ie_i$, and $$v\wedge
d_Av\equiv\sum_{i,j=1}^5v_i (d_Av)_j[e_i\wedge e_j]=\sum_{i,j=1}^5v_i
(d_Av)_jE_{ij}.$$ We have the
following properties
\begin{itemize}
\item
$\omega$ takes values in $\mathfrak{so}(5)$.

\item $\omega$ vanishes on vertical vectors.
\item For $g\in G,p\in P, X\in T_pP$ we have
$\omega((R_g)_*X)_{pg}=\text{ad}_{g^{-1}}\omega(X)_p$.

\end{itemize}
Therefore $\omega\in\Omega^1(W,\text{ad} P)$.
We can therefore consider the connection on $P$

$$B=A-\omega=A-v\wedge d_Av$$
Assume now that $\langle v,v\rangle=1$.
Then
$$d_Bv=dv-(A-v\wedge d_Av)v=dv-Av+(v\wedge d_Av)v=dv-Av-d_Av=0$$

In fact
\[{[v_i d_Av_jv_k(E_{ij})e_k]}_m=v_i d_Av_j v_k
{[E_{ij}]}_{ml}{(e_k)}_l=\] \[
=
v_m d_Av_j v_j- v_k d_Av_m v_k  =
(\langle d_Av,v \rangle v-\langle v,v\rangle d_Av)_m=-d_Av\]
because $\langle d_Av,v\rangle=0$  if we assume that $\langle v,v\rangle=1$.
The equation $d_B \v=0$, means that
$v$ is parallel with respect to $B$, i.e. $B$ {\it is a reducible connection}
(in $P$), with reduced group $SO(4)$ (see \cite{KN}, vol.1, Proposition 7.4 of
Chapter 2).
$B$ is therefore a  connection reducible to the subbundle determined by $v$.
 We will make now the
assumption that the bundle $P$ has a double covering $\widetilde P\to W$
with structure group
$Spin(5)$.
In order for this assumption to hold it is enough that both $W$ and $X$
have a spin-structure.
Now every connection $A$ on $\widetilde P$ induces a connection, whih we
call again $A$ on $P$ via
the isomorphism \ref{iso}. The above argument generalizes to $\widetilde P$.

If, however, one works in $\widetilde P$ with the spin connection
and use the identifications described above, one must be
aware that many familiar conventions change. In the next subsection we collect
a set of useful formulas and results by making use of this formalism. 

\subsection{Some explicit formulae and the Pfaffian}

For simplicity, let us speak about a principal fiber bundle $P$ with structure
group $SO(5)$, a connection $A$ in it and a map $v$ from $W$ to $\mathbb R^5$.
We start with 
\be
&& d_A\v = d\v - \frac{1}{2}[A,\v]\nonumber\\
&& F_A= dA - \frac{1}{4} [A,A]\nonumber\\
&& d_Ad_A \v = -\frac{1}{2}[F_A,\v]\nonumber
\ee
The transformation properties under a gauge transformation
$\Lambda=\Lambda^{ab}\Sigma_{ab}$ are
\be
&&\delta \v = -\frac{1}{2} [\v , \Lambda]\longrightarrow
\delta v^a = \Lambda^a{}_b v^b\nonumber\\
&&\delta A = d_A\Lambda= dA - \frac{1}{2} [A, \Lambda]\nonumber\\
&& \delta F_A = - \frac{1}{2} [F,\Lambda]\nonumber
\ee
and so on.
Notice that we have
$$
\sum_{a=1}^5 {\rm Tr}(\gamma_a \Sigma_{j_1j_2}\Sigma_{j_3j_4})
{\rm Tr}(\gamma^a \Sigma^{i_1i_2}\Sigma^{i_3i_4})= 5\Big(\delta_{j_1}^{i_1}
\delta_{j_2}^{i_2}\delta_{j_3}^{i_3}\delta_{j_4}^{i_4}\pm {\rm permutations}
\Big)
$$
and so
\be
p_2(N)= \frac{1}{(2\pi)^4 4!} \sum_{a=1}^5 {\rm Tr}(\gamma_a FF) {\rm Tr}
(\gamma^a FF),
\quad\quad F= F^{ab}\Sigma_{ab}\label{p2N}
\ee
Setting
$$
\chi_a = {\rm Tr}(\gamma_a FF)=\epsilon_{abcde}F^{bc}F^{de}
$$
and $\Lambda=\Lambda^{ab}\Sigma_{ab}$, we get
$$
\delta \chi_a = -2 \Lambda_a{}^b \chi_b
$$

We will often use the following obvious {\it vanishing argument}. Write
$$
\epsilon_{abcde}v_1^av_2^bv_3^cv_4^dv_5^e \sim {\rm Tr} (\v_1 \v_2
\v_3 \v_4 \v_5)
$$
where $v_i$, $i=1,\ldots,5$ are vectors in  
${\mathbb R}^5$. If they are all orthogonal to the same vector $v$ in
then the above expression vanishes.

The reducible connection $B$ in $P$ becomes in $\widetilde P$
\be
B_v = A -\frac{1}{2} [\v, d_A \v]\label{redconn}
\ee
Very often we will drop the label $v$ in $B_v$. Whenever this happens it is
understood that we refer to the reduction represented by the $v$ in question.
We will not insist either on the distinction between $P$ and $\widetilde P$.
It is easy to directly verify that
\be
d_{B_v}\v\equiv d_B \v =0\label{dBv=0}
\ee
where use has been made of $\langle v,d_Av\rangle =0$. As a 
consequence of (\ref{redconn}) we have:
$$
F_B = F_A + \frac{1}{4}[\v,[F_A, \v]] - \frac{1}{4} [d_A \v,
d_A\v]
$$
From (\ref{dBv=0}) we also deduce that $[F_B,\v]=0$. This in turn implies that
$F_B\perp v$, i.e. $F_B^{ab}v_a=0 = F_B^{ab}v_b$.

Another useful result is the following: if
$Z\in aut P$ is  vertical, then, from (\ref{redconn}), $B(Z)=A(Z)$.

Let us consider next the effect of a generic $Z\in aut P$ on $v$. We have
\be
L_Z \v= \frac 12 [i_ZB, \v]\label{LZv}
\ee
since $[i_Z F_B, \v]=0$.
Any transformation $\delta v$ of this kind maps {\it a reducible connection
$B$ into a new reducible connection $B+ L_Z B$}. In fact
$$
d_{B+L_Z B} (\v+ L_Z \v) =0
$$
up to infinitesimals of higher order. Notice that
$Z$ maps $N$ to $N$, but deforms $N'_v$, for it maps it to $N'_{v+\delta v}$.

Any reduction $R$ of $P$ determines a factorization of $p_2(N)$ into the
square of $Pf(A,v)={\rm Tr}(\v F_BF_B)$. In fact, given a reduction
determined by
a vector $v$, it makes sense to consider in $p_2(N)$ only the relevant
reducible connections.
Therefore in (\ref{p2N}) we must replace everywhere $A$ with $B\equiv B_v$.
But since, as we have noticed, $F_B \perp v$, only the component of $\chi_a$
parallel to $v$ will contribute in the expression of $p_2(N)$ due to the
vanishing argument. Therefore we can write
\be
\left. \frac{1}{24}p_2(N)\right\vert_v =
\alpha^2\Big(Pf(A,v)\Big)^2\label{p2N'},
\quad\quad Pf(A,v)= {\rm Tr} (vF_BF_B),\quad\quad
\alpha = (24)^{-1}(2\pi)^{-2}\label{split}
\ee
with obvious meaning of the subscript $v$. The number of independent
degrees of freedom in $v$ is 4, the same as the difference between the $SO(5)$
gauge variables and the $SO(4)$ ones. What happens is clear: when considering
reducible connections we trade the gauge parameters lost in the reduction
process with the free parameters in $v$.

Notice that in $Pf(A,v)$, the Pfaffian corresponding to $N_v'$, only terms
linear in $\v$ and with an even number of $d_A\v$ survive (compare with
\cite{FHMM}).

\section{Expression of the anomaly}

Our aim is to deduce the anomaly from the usual descent equations, starting
from the Pfaffian $Pf(A,v)$ introduced above
while introducing a background connection.
The derivation is far from straightforward. So before we deal with it, let us
work out a simpler well--known example and use it as a guide.

Let us consider a generic connection $A$ in a principal bundle $P$ with
base $M$, together
with a background connection $A_0$. Let $I$ denote
the unit interval over which a parameter $t$ is defined. Then we can think
of the interpolating connection ${\cal A}_t=A_0+t(A-A_0)$ as a connection on
$P\times I$ and denote it $\hat {\cal A}$ (from now on hatted symbols will
indicate quantities relevant to $P\times I$). Its curvature will be
$$\hat{\cal F}={\cal F}_{{\cal A}_t}-(A-A_0)dt$$
Correspondingly, if $d$ denotes
the exterior derivative in $M$, $\hat d$ will be the exterior derivative on
$M\times I$. We can easily derive the Chern--Simons formula:
$$d\int_I Q(\hat{\cal F} )=\int_I \hat dQ(\hat{\cal F})-\left.Q({\cal
F}_t)\right\vert^1_0=-Q(F)+Q(F_0)$$
where $Q$ is any symmetric ad--invariant polynomial\footnote{$Q$ has $n$
entries, but here and in the following we adopt the convention that whenever
several entries coincide, we write down only one of them,
say $Q(F,\ldots,F)\equiv Q(F)$.}.
So the Chern--Simons term can be written $$W_Q(A,A_0)=-\int_I Q(\hat{\cal F} )$$
We would like now to express in a similar way the corresponding anomaly.
Consider the derivative in the space of connections on $P\times I$ along
 the vector $t\xi$, with $\xi=L_ZA\in \Omega^1(M,\text{\rm ad}P)$,
 for any $Z\in autP$.
Then $\delta \hat{\cal F}=\hat d_{\hat{\cal A}}t\xi=td_{{\cal A}_t}\xi-\xi
dt$,
and
$$
\delta \int_I Q(\hat{\cal F})=\int_I n Q(\delta\hat{\cal F},\hat{\cal F})=
\int_I n Q(\hat d_{\hat{\cal A}}t\xi,\hat{\cal F})=
\int_I n \hat dQ( t\xi,\hat{\cal F})=$$ $$
d\int_I   nQ( t\xi,\hat{\cal F})- nQ(\xi,F)=d\left(n(1-n)\int_Idt~  tQ(
\xi,A-A_0,{\cal F}_{{\cal A}_t})\right)-nQ(\xi,F)
$$
The last term drops for dimensional reasons (if $dim~M\leq 2n-2$) and the 
term in brackets gives the well--known expression of the anomaly.

Let us return now to our original problem on the 5--brane world volume $W$.
For a consistent construction we have to introduce both a reducible reference
connection $A_0$ and a corresponding reference reduction $v_0$.
So we can define $Pf(A_0,v_0)= {\rm Tr}(\v_0 F_{B_0}F_{B_0})$.
Now we introduce the interpolating connection ${\cal A}_t = tA + (1-t)A_0$
and a path of unit vectors $v_t$ that interpolates between $v_0$ and $v_1=v$.
There are of course many different paths from $v_0$ to $v$. All we say in this
and the following section does not depend on what path we choose. However
this issue will become important in the final section, in which we will need
the path to depend only on the initial and final vectors. To satisfy this
requirement we select one single path from $v_0$ to $v=v_1$: since $v_0$
and $v$
lie both on a sphere $S^4$ embedded in ${\mathbb R}^5$, we will choose the
geodesic (with respect to the embedded metric) passing through $v_0$ and $v$
\footnote{This prescription is not unique whenever $v_1=-v_0$. However, except
when both $v_0$ and $v_1$ are constant sections, a case we can easily exclude,
the ambiguity can be resolved by continuity.}.
This is a canonical choice of a path between $v_0$ and $v$.

Now we set
$$
{\cal B}_{v_t}\equiv {\cal B}_t =
{\cal A}_t - \frac{1}{2} [\v_t, d_{{\cal A}_t} \v_t]
$$
We have, of course,
$$
d_{{\cal B}_t} \v_t=0
$$

Now let us consider the family $R_t$ of reductions of $P$, determined by
$v_t$.
Then we have a bundle ${\cal R}\to W\times I$. The path ${\cal B}_t$,
just introduced, can be viewed as a connection $\cal B$ on ${\cal R}$.
We can then repeat the above procedure.
For simplicity, from now on we write our Pfaffian in a more standard way:
$$Q(A,B,C)\equiv \alpha {\rm Tr} (ABC)$$
(the constant $\alpha$ has been introduced in (\ref{split})).
The polynomial $Q$ is indeed symmetric and ad--invariant.

Let ${\hat {\cal A}}\equiv {\cal A}_t=A_0+t(A-A_0)$ on
$P\times I$ and $\hat{\cal B}={\cal B}_t-\frac 12[\v_t,\dot \v_t]dt=
{\cal A}_t-\frac
12[\v_t,\hat d_{\hat {\cal A}}\v_t]$.
Then $\hat d_{\hat{\cal B}}\v_t=d_{{\cal B}_t}\v_t+
\left(\dot \v_t+\frac 14[ [\v_t,\dot
\v_t],\v_t]\right)dt=0$ (using the gamma matrix algebra)
and $\hat d_{\hat{\cal B}} \hat{\cal F}_{\hat{\cal B}}=0$, where
$$\hat{\cal F}_{\hat{\cal B}}={\cal F}_{{\cal B}_t}-\left(\dot {\cal B}_t
+\frac 12 d_{{\cal B}_t}([\v_t,\dot\v_t])\right)dt$$

Correspondingly we have
$$d\int_I Q(\v_t,\hat{\cal F}_{\hat{\cal B}},\hat{\cal F}_{\hat{\cal B}}
)=\int_I
\hat dQ(\v_t, \hat{\cal F}_{\hat{\cal B}}
,\hat{\cal F}_{\hat{\cal B}})-\left.Q(\v_t,{\cal F}_{{\cal B}_t} ,
{\cal F}_{{\cal B}_t})
\right\vert^1_0=-Q(\v_1,F_B,F_B)+Q(\v_0,F_{B_0},F_{B_0})
$$
So the Chern--Simons term we were looking for is
\be
W_Q(v,v_0,B,B_0)=-
\int_I  Q(\v_t,\hat{\cal F}_{\hat{\cal B}},\hat{\cal F}_{\hat{\cal B}})
\label{CS}
\ee

We would like now to express in a similar way the anomaly. Consider the
derivative in the space of connections on $\cal R$ along
the vector $t\xi$, with $\xi=L_ZA\in \Omega^1(W,\text{\rm ad}P)$ {\it and}
along $\delta \v= L_Z\v$. Then
$$
\delta \hat{\cal B}\equiv \delta ({\cal B}_t- \frac 12 [\v_t,\dot \v_t] dt) =
\hat G(\xi)
$$
where
\be
\hat G(\xi) =t(\xi +\frac 14 [\v_t,[\xi,\v_t]]) -
\frac 12 [\delta \v_t , \hat d_{{\cal A}_t} \v_t]
- \frac 12 [\v_t, \hat d_{{\cal A}_t}\delta \v_t]\equiv G_t(\xi)+(\ldots)dt\0
\ee
and $\hat d_{{\cal A}_t}\v_t = d_{{\cal A}_t} \v_t + \dot \v_t dt,$ etc.
Then
$$
\delta \hat{\cal F}_{\hat{\cal B}}= \hat d_{\hat{\cal B}}\hat G(\xi)
$$

We have
\be
\delta \int_I Q(\v_t,\hat{\cal F}_{\hat{\cal B}},\hat{\cal F}_{\hat{\cal
B}})&=&
\int_I Q(\delta \v_t, \hat{\cal F}_{\hat{\cal B}}, \hat{\cal F}_{\hat{\cal
B}})
+ 2\int_I
Q(\v_t,\delta\hat{\cal F}_{\hat{\cal B}},\hat{\cal F}_{\hat{\cal B}})\0\\
&=&2\int_I  Q(\v_t,\hat d_{\hat{\cal B}}\hat G(\xi),\hat{\cal F}_{\hat{\cal
B}})=
2\int_I   \hat d Q(\v_t, \hat G(\xi),\hat{\cal F} _{\cal B})\0\\
&=&2d\left(\int_I  Q(\v_t,\hat G(\xi),\hat{\cal F}_{\hat{\cal B}})
 \right)-2Q(\v_1,G_1(\xi),F_B)\label{anom1}
\ee
In fact
\be
\int_I Q(\delta \v_t, \hat{\cal F}_{\hat{\cal B}}, \hat{\cal F}_{\hat{\cal
B}})&=&
2 \int dt Q(\delta \v_t, \dot {\cal B}_t+
\frac 12 d_{{\cal B}_t}[\v_t, \dot \v_t],
{\cal F}_{{\cal B}_t})\0\\
&=& 2 \int dt Q(\delta \v_t, \dot {\cal B}_t+\frac 14 [\v_t,[\dot {\cal B}_t,
\v_t]], {\cal F}_{{\cal B}_t})\label{vanish}
\ee
having used $d_{{\cal B}_t}\v_t=0$ differentiated with respect to $t$. Now
it is
easy to prove that any quantity
$$
\hat C = C + \frac 14 [\v,[C,\v]],\quad\quad C= C^{ab}\Sigma_{ab}
$$
is such that $[\hat C, \v]=0$, i.e. $\hat C \perp \v$. Therefore, using
(\ref{LZv}) applied to $\v_t$ and using the ad--invariance of the polynomial
$Q$, it is easy to see that (\ref{vanish}) vanishes.

Now we recall that
\be
\delta v= L_Z v, \quad\quad \delta v_0=0\label{assume}
\ee
and notice that $G_1(\xi) = L_Z B$. Therefore we can write
\be
\delta \int_I Q(\v_t,\hat{\cal F}_{\hat{\cal B}},\hat{\cal F}_{\hat{\cal
B}})&=&
2d\left(\int_I  Q(\v_t,\hat G(\xi),\hat{\cal F}_{\hat{\cal B}}) -
Q(\v, i_ZB, F_B) \right)\0\\
 &-& i_Z Q(\v,F_B,F_B) \0\\
&\equiv& d {\cal A}_2^1 - i_Z Q(\v,F_B,F_B)\label{anom'}
\ee
${\cal A}_2^1$ corresponds to the familiar 2d anomaly.
The expression of ${\cal A}_2^1$ is as follows
\be
{\cal A}_2^1&=&  -2 \Big(\int dt Q(\v_t,L_Z {\cal B}_t, \dot
{\cal B}_t+\frac 14 [\v_t,[\dot {\cal B}_t,\v_t]]) + Q(\v, i_ZB, F_B)\Big)\0\\
&-&\int dt Q(\v_t,[L_Z \v_t,\dot \v_t], {\cal F}_{{\cal B}_t})
- \int dt Q(\v_t,[\v_t, \delta \dot \v_t], {\cal F}_{{\cal B}_t})\label{A21}
\ee
The last term vanishes since $[{\cal F}_{{\cal B}_t},\v_t]=0$.

Now let us extract the full anomaly. The Chern--Simons term is
$$
W_{tot}(\v,\v_0,B,B_0)= W_Q(\v,\v_0,B,B_0) \Big(Q(F_B)+ Q(F_{B_0})\Big)
$$
where $Q(F_B)= Q(\v, F_B,F_B)$, $Q(F_{B_0})= Q(\v_0, F_{B_0},F_{B_0})$,
and
$$
 W_Q(\v,\v_0,B,B_0)= - 2 \int dt Q(\v_t, \dot {\cal B}_t+
\frac 14 [\v_t,[\dot {\cal B}_t,
\v_t]], {\cal F}_{{\cal B}_t})
$$
With standard steps we obtain
\be
\delta W_{tot} = i_Z (Q(F_B) Q(F_B))-d \Big[{\cal
A}_2^1\left(Q(F_B)+Q(F_{B_0})\right)+W_Q(\v,\v_0,B,B_0) i_Z Q(F_B) \Big] \0
\ee
The first term on the RHS vanishes for dimensional reasons. The term in square
bracket is the anomaly.
Therefore
\be
Anom_v = {\cal A}_2^1 (Q(F_B)+Q(F_{B_0})) +
W_Q(\v,\v_0,B,B_0) i_Z Q(F_B)\label{Anomv}
\ee
The label $v$ is in order to stress that this expression depends on the
particular reduction $v$ we have chosen.

\section{The M--5--brane anomaly cancellation}

Before we embark in the discussion of anomaly cancellation, let us summarize
what we have done and
clarify the role of the background connection $A_0$ in this
context. The anomaly just obtained is
the residual M--5--brane anomaly generated via the descent equations from
the 8--form $p_2(N)$, i.e. the second Pontryagin class of the normal bundle.
For reducible connections $p_2(N)$ splits into the square
of a Pfaffian.  Any such splitting can be thought to correspond
to a section $v$. Therefore we can write
\be
N= L_v\oplus N'_v\label{normsplit}
\ee
At this point we remark that, if in $p_2(N)$ we consider a connection reducible
to $SO(4)$ (i.e. the connection $B_v$ introduced above),
the second Pontryagin class of $N$ becomes:
\be
p_2(N)= p_2(N'_v)= e(N'_v)^2= 24~ Q(\v, F_B,F_B)^2\label{e2'}
\ee
where $e$ represents the Euler class. Therefore the anomaly
we have to cancel is generated by $Q(\v,F_B,F_B)$ and given by (\ref{Anomv}).

It is thanks to the factorization (\ref{e2'}) that in the previous section
we were able to derive eq.(\ref{Anomv}) for the anomaly, and in this
section we are able to cancel it via a suitable counterterm.
At this point however, as partially anticipated in the previous section,
we can take two different attitudes.

The first attitude is based on the idea that one has
to restrict the subgroup of relevant diffeomorphisms of $X$ to those
which, not only map $W\to W$ and $N\to N$, but also preserve the splitting
$N= L_v \oplus N_v'$, i.e. in particular preserve $v$. Let us call this
subgroup $Diff_v(X,N)$. Now we can safely pick a background connection
$A_0$ in $N'_v$: it will remain a connection in any transformed $N'_v$, as
long as the diffeomorphisms considered are those of $Diff_v(X,N)$.
They correspond to the automorphisms of a principal fiber bundle $P'$ whose
structure group is $SO(4)$. With this understanding the anomaly
is given by (\ref{Anomv}) with fixed $v$. This attitude
assumes that $v$ has some physical meaning.
We recall that $L_v$ represents the direction normal to $W$ which
is tangent to $Y$, the manifold which bounds $W$. This means therefore
that $Y$, or at least a collar which represents the part of $Y$ nearest to $W$,
retains some physical information too. In other words, it would
seem that physical information about the 5--brane is not stored only in
$W$ but also on $Y$. This sounds curiously similar to what has been proposed
in a different context in \cite{wittenads}, where two different manifolds
terminating on the same space--time manifold carry different physical
information. We shall call this scheme the {\it restricted scheme}.
This is essentially the scheme adopted in \cite{BCR,BR}. Another context in
which
this scheme applies is to describe the compactification of M theory on a
circle $S^1$ along a direction transverse to the 5--brane. In this case the
latter becomes the NS 5--brane of type IIA theory. As pointed out in
\cite{witten}, the normal bundle to the 5--brane then splits into the direct
sum of a vector bundle with structure group $SO(4)$ and a trivial line bundle,
which actually coincides with the tangent bundle of the compactification
circle.
Although the construction is not exactly the same as above, the final setup
is. Therefore we can choose a section of the trivial line bundle, say $v$, and
redo everything without changing a single word. In this case the physical
nature
of $v$ is immediately visible.

The second attitude, or {\it general scheme}, assumes instead that, although
the splitting $N= L_v \oplus N'_v$ has of course a physical meaning, since it
represents, via (\ref{df4}), the magnetic coupling of the 5--brane to M theory,
there is no privileged reductions.
Therefore we have to consider all possible $v$'s and integrate over them in the relevant path integral
(see below for further comments on this point).
In this case the gauge transformations are all $Z\in aut P$, where $P$ is the
principal fiber bundle with structure group $SO(5)$ associated to $N$.
The intermediate steps in the derivation of the anomaly make sense since
we have seen that a
transformation by any $Z\in aut P$ maps a reduction to another reduction and a
reducible connection into another
reducible connection. There is no problem either with the background
connection $A_0$
and the background vector $v_0$, which remain fixed throughout. Finally the
anomaly is again given by (\ref{Anomv}).

\subsection{The counterterm}

The basis for anomaly cancellation is that $e$ be cohomologically trivial,
i.e.
\be
Q(\v,F_B,F_B)= \left. d\eta_v\right\vert_W\label{triv}
\ee
where $\eta_v$ is some 3--form field in the theory. We will discuss in the
following
subsection how to relate $\eta_v$ to the theory. For the time being let us
suppose that it exists and is local.

Consistently with (\ref{triv}) we set
\be
Q(\v_0,F_0,F_0)= d\eta_{v_0} \label{triv'}
\ee
where $\eta_{v_0}$ need not be a local field, it may be a purely
differential--geometric 3--form. Moreover, consistently with our
definitions and with $\delta A_0=0$, we set
\be
\delta \eta_v = L_Z \eta_v, \quad\quad  \delta \eta_{v_0} =0\label{deltaeta}
\ee

As we said above, the anomaly we have to cancel is given by (\ref{Anomv}).
Our proposed counterterm to cancel it is
\be
S_v=\int_W \Big((\eta_v+\eta_{v_0})\wedge W_Q(\v,\v_0,B,B_0)+
\eta_v\wedge \eta_{v_0}
\Big)
\label{counter5}
\ee

We have
\be
\delta S_v &=& \int_W \Big[L_Z\eta_v \wedge  W_Q(\v,\v_0,B,B_0) +
L_Z \eta_v \wedge\eta_{v_0} \0\\
&&~~- (\eta_v+\eta_{v_0})\wedge \Big( d {\cal A}_2^1 - i_Z
Q(F_B)\Big)\Big]\0\\
&=&\int_W \Big[ i_Z Q(F_B) \Big( W_Q(\v,\v_0,B,B_0) +\eta_{v_0}\Big)- i_Z\eta_v
\wedge
d\Big( W_Q(\v,\v_0,B,B_0)+ \eta_{v_0}\Big)\0\\
&&~~+ (\eta_v+\eta_{v_0})\wedge i_ZQ(F_B)
- (Q(F_B)+Q(F_{B_0})) {\cal A}_2^1 \Big]\0\\
&=&\int_W \Big[ i_Z Q(F_B)  W_Q(\v,\v_0,B,B_0) - i_Z(Q(F_B)\wedge
\eta_v)\0\\
&&~~ -(Q(F_B)+Q(F_{B_0})) {\cal A}_2^1 \Big]\0\\
&=& \int_W\Big[i_Z Q(F_B) W_Q(\v,\v_0,B,B_0)-
(Q(F_B)+Q(F_{B_0})) {\cal A}_2^1 \Big]\0
\ee
where we have used $i_Z[\eta_v\wedge Q(F_B,F_B)]=0$ for dimensional reasons.
Therefore adding the counterterm $S$ to the action cancels
the residual M--5--brane anomaly.

The anomaly cancellation works in both restricted and general schemes.

\subsection{The nature of $\eta_v$}

The previous cancellation mechanism is based on the existence in the
M--theory with a 5--brane of a 3--form field with the transformation
property (\ref{df4}). In such a theory there are several 3--forms.
From 11 dimensional supergravity we have the 3--form $C_3$. On the 5--brane
we have a 2--form $B_2$, by means of which in the interacting theory we
can form the combination $H_3=dB_2 - C_3$. But neither $C_3$ nor $H_3$ can
be identified with $\eta_v$, even though the transformation property would
be the right one (\ref{deltaeta}): we know that in the absence of
the 5--brane we have $dC_3= F_4$, while, when the 5--brane is present,
$F_4$ is modified in such a way that (\ref{df4}) holds; therefore $C_3$
does not contain any information concerning $e$.

From this discussion it is evident that $\eta_v$ must be constructed out of
$F_4$. Let us generalize the construction presented in \cite{BCR}.
The section $v$, which determines the decomposition $N= L_v\oplus N_v'$,
can be seen as a vector field on the 11 dimensional space $X$:
${\mathfrak v} = \sum_i {\mathfrak v}^i \frac {\d}{\d x^i}$. In fact $v$
is a section of a line bundle which lies in $N$ and therefore in $TX$
\footnote{We recall
that $\mathfrak v$ is a vector field in $X$, while $v$ represents a set
of scalar fields in $W$}. In the following we will need the equation
\be
\left. i_{\mathfrak v}\Phi(L)\right\vert_W=1\label{norm}
\ee
where $\Phi$ denotes the Thom form on $L$. To show (\ref{norm}),
notice that for a generic
vector field ${\mathfrak v}$ with nonzero components only along $L$,
$\left. i_{\mathfrak v}\Phi(L)\right\vert_W$ is
a non--vanishing function on $W$, therefore a suitable rescaling is enough to
produce the desired result. Whichever the choice, we remark that what we
have achieved so far is the definition of
${\mathfrak v}$ only on $W$; outside $W$ we can define it in an
arbitrary way.

Now, assuming the triviality of $Q(\v,F_B,F_B)$, eq.(\ref{df4}), given
${\mathfrak v}$, is equivalent to the existence
of a local 3--form $\chi_v$ that solves the
equation
$$
\left. L_{\mathfrak v}F_4\right\vert_W=d\chi_v
$$
In fact using $L_{\mathfrak v}=di_{\mathfrak v}+i_{\mathfrak v}d$ we have
$$
\left. \left(di_{\mathfrak v}F_4+i_{\mathfrak
v}dF_4\right)\right\vert_W=d\chi\Leftrightarrow
\left. i_{\mathfrak v}\delta_W
\right\vert_W=d_W(\chi_v-\left. i_{\mathfrak v}F_4\right\vert_W)
$$
We recall from \cite{BCR} that $\delta_W=\Phi(N_v')\wedge\Phi(L_v)$ and
the Thom form $\Phi(N_v')$ can be interchanged with the Euler form $e$ of
$N_v'$.
Then
we get $$\left. i_{\mathfrak v}\delta_W\right\vert_W=\left.
\Phi(N_v')\right\vert_W=e(N_v')=
Q(\v,F_B,F_B)$$
So, finally,
$$Q(\v,F_B,F_B)=\left. d\eta_v\right\vert_W, \quad\quad \eta_v= \chi_v -
i_{\mathfrak v} F_4$$
This $\eta_v$ satisfies all the requirements. The form $\chi_v$ is left 
undetermined by our analysis.

\subsection{The fate of $v$}

It remains for us to discuss the implications that come for a theory from
the presence of $v$. In the general scheme, in fact, we have still to explain
how we deal with $v$ from a field--theoretic point of 
view: what kind of field is $v$ and what is the path integral treatment of it
beyond the anomaly problem, where $v$ is a spectator?

We summarize what has been said so far: the possible magnetic couplings
of the 5--brane inside the M--theory are spanned by the $v$ sections. Each
$v$ represents a reduction from the structure group of the normal
bundle of $W$ from $SO(5)$ to $SO(4)$. In view of this physical input (i.e. the
magnetic coupling) it only makes sense to consider reducible connections,
i.e. connections valued in the Lie algebra ${\mathfrak{so}}(5)$, which, when 
restricted to the reduced bundle, are connections valued in the Lie 
subalgebra ${\mathfrak {so}}(4)$. Given a connection $A$ in the principal 
fiber bundle $P$ with gauge group
$SO(5)$, and a section $v$ of the associated bundle with fiber $\frac{SO(5)}
{SO(4)}$, we can construct a reducible connection $B_v$ via (\ref{redconn}).
Therefore the relevant theory is obtained starting from the theory
in the eleven dimensional manifold $X$ coupled to the 5--brane with
world--volume $W$, considering the splitting of the spin 
connection of $X$ on $W$ into a tangential and normal part (the latter is 
exactly $A$) and then replacing $A$ everywhere with $B_v$. Here we have to be
careful about possible Jacobian factors. The measure over $A$ in the path 
integral is provided by the theory. As for the measure over $v$, it is very 
natural to adopt the measure of the gauge transformations that map $v$ to a  
fixed
section $v_0$ (we have already remarked that, in the process of reducing the 
structure group, we trade such gauge transformations for the $v$'s). Now,
surprisingly enough, the Jacobian for the passage from $A$ to $B_v$ at fixed
$v$ is a constant. Therefore we can use, as path integral measure for the
theory formulated in terms of $B_v$, the product of the measure of $A$ and
the measure of $v$.  

After defining the relevant path integral measures, let us turn to the 
action. Now comes the crucial point: as remarked above, although the action is 
now expressed in terms of
reducible connections, the gauge symmetry group is still $SO(5)$ (not simply
$SO(4)$), because an $aut P$ transformation (with structure group $SO(5)$) maps a 
reducible connections into reducible connections while keeping the the
background connection unchanged -- we stress that
this is true in the present case, but is not true in general. This is reflected
in the fact that we have considered anomalies of $aut P$, not anomalies of the
gauge transformations with group $SO(4)$. 

Next, while computing anomalies, both $A$ and $v$ are spectators. We have seen
that a suitable choice of the counterterm (depending on $A$ and $v$) allows us 
to free the theory from anomalies. The question is now: what do we do next with
$B_v$ and $v$? In particular, what is the fate of $v$? 

If we go on with the path integral quantization, after taking care of the 
fermions determinants, it is necessary to fix the gauge. A simple
way (not necessarily the best one) to do it is the following. The infinitesimal 
gauge parameters \footnote{For global gauge transformations it is perhaps 
necessary to deal with this problem more carefully.}
split into the direct sums of gauge transformations that leave $v$ invariant 
and the ones that modify $v$. We can fix the gauge by first choosing one fixed
$v$, say $v_0$, and then fixing the remaining $SO(4)$ (or, better, $aut P'$ with
structure group $SO(4)$) gauge invariance in the
ordinary Faddeev--Popov way\footnote{It would seem that in this way we return 
to the restricted scheme. This is not so, because we first make sure that
anomaly are canceled so that the full gauge symmetry is restored, and only 
afterwards do we fix the gauge.}.

In this way we have closed the circle. What was originally a set of gauge 
degrees of freedom (i.e., $v$) have met their fate, that is they have been 
gauge--fixed and have disappeared from the game (except for the remnant $v_0$).

To conclude, it is worth emphasizing the roots of the successful 
cancellation of the M--5--brane anomaly, which originates from the fundamental 
role played by eq. (\ref{df4}). This equation entails the normal bundle 
splitting, which, in 
turn, implies that the second Pontryagin class of the normal bundle is 
factorizable as the square of the Pfaffian. It is only thanks to such occurrence
that we can write down the local counterterm (\ref{counter5}).

\subsection{One final comment}

Our approach in this paper is partially on--shell. In fact we suppose
throughout that (\ref{df4}) be satisfied. It would be interesting to know
whether one can extend it to an off--shell treatment.
In \cite{PST} a local action for the M--5--brane was proposed that overcomes the
traditional difficulty connected with the kinetic term of the self--dual
two form of the M--5--brane. This is done by introducing
additional fields and gauge symmetries so that the additional degrees of
freedom turn out to be pure gauge and the gauge freedom implies the sought for
equations of motion. In \cite{BBS} this scheme was extended by embedding
the M--5--brane action into the full action of 11 dimensional supergravity.

We do not know whether one can deal with the anomaly problem in such more
general framework. However it is interesting to point out some similarities. 
One of the additional fields introduced in \cite{BBS} is an eleven
dimensional vector field $u$, a section of $TX$. In the presence of the
5--brane $u$, restricted to $W$, splits naturally into two components, one
in the tangent bundle to $W$ and another in the normal bundle. It is natural to
identify the latter components with the section $v$ which is responsible for
the reduction of the structure group of the normal bundle. Said otherwise,
we can immerse ${\mathfrak v}$ in the formalism of \cite{BBS} by assimilating
it to the additional $u$ field. The latter is then dealt with as a pure
gauge degree of freedom.  

\vskip1cm

{\bf Acknowledgements} One of us (L.B.) would like to thank P.Pasti, D.Sorokin
and M.Tonin for very stimulating discussions. We would like to thank C.S.Chu
for his collaboration in an early stage of this work. 
This work was partially supported by EC TMR Programme,
grant FMRX-CT96-0012, and by the Italian MURST for the program
``Fisica Teorica delle Interazioni Fondamentali''.


\begin{thebibliography}{99}

\bibitem{KN} S.Kobayashi, K.Nomizu {\it Foundation of Differential Geometry}
vol. I, New York, London. Interscience 1963,1969.

\bibitem{MSZ} J.Ma$\tilde{\rm n}$es, R.Stora, B.Zumino, {\it Algebraic study
of chiral anomalies}, Comm.Math.Phys. 102 (1985) 157.

\bibitem{BCRS1} L. Bonora, P. Cotta--Ramusino, M. Rinaldi and J. Stasheff,
{\it The Evaluation Map in Field Theory, Sigma--Models and Strings I},
Comm. Math. Phys. 112 (1987) 237.

\bibitem{witten} E. Witten, {\it Five-Brane Effective Action in M--theory},
[\hepth{9610234}].

\bibitem{alwis} S.P.de Alwis, {\it Coupling of Branes and Normalization of
Effective Actions in String/M--theory}, Phys.Rev. D56 (1997) 7963-7977,
[\hepth{9705139}]

\bibitem{henningson} M.Henningson, {\it Global anomalies in M--theory}
Nucl.Phys. B515 (1998) 233-245.

\bibitem{BCR} L.Bonora, C.S.Chu, M.Rinaldi, {\it Perturbative Anomaly of the
M--5--brane}, JHEP 12 (1997) 007, [\hepth{9710063}].

{\it Anomalies and locality in field theory and M theory}
[\hepth{9712205}], Proc. Moscow Conference on Secondary Calculus and
Cohomological Physics. AMS 219.

\bibitem{BR} L.Bonora, M.Rinaldi, {\it Branes, Normal Bundles and Anomalies},
to appear in a volume in honor of C.Becchi.

\bibitem{FHMM} D.Freed, J.A.Harvey, R.Minasian, G.Moore, {\it Gravitational
Anomaly Cancellation for M--Theory Fivebranes}, Adv.Theor.Math.Phys. 2
(1998) 601,
 [\hepth{9803205}].

\bibitem{BB} K.Becker and M.Becker, {\it Five--brane Gravitational Anomalies}
[\hepth{9911138}].

\bibitem{mourad} J.Mourad, {\it Anomalies of the SO(32) five-brane and
their cancellation}, Nucl.Phys. B512 (1998) 199-208, [\hepth{9709012}].

\bibitem{braxmourad} Ph.Brax, J.Mourad {\it Open Supermembranes Coupled to
M--Theory Five--Branes}, Phys.Lett. B416 (1998) 295-302, [\hepth{9707246}].


\bibitem{CY} Y.-K. E.Cheung, Z.Yin, {\it Anomalies, Branes, and Currents}
Nucl.Phys. B517 (1998) 69-91, [\hepth{9710206}].

\bibitem{wittenads} E.Witten {\it Anti De Sitter Space And Holography},
Adv.Theor.Math.Phys. 2 (1998) 253-291, [\hepth{9802150}].
{\it Anti-de Sitter Space, Thermal Phase Transition, And Confinement In Gauge
Theories}, Adv.Theor.Math.Phys. 2 (1998) 505-532, [\hepth{9803131}].

\bibitem{PST} P.Pasti, D.Sorokin and M.Tonin,{\it On Lorentz invariant actions
for chiral
p--forms}, Phys.Rev. D52 (1995) 4277,[\hepth{9611100}];
{\it Covariant action for a D=11 5--brane with the chiral field}, Phys.Rev. D55
(1977) 6292, [\hepth{}].

\bibitem{BBS} I.Bandos, N.Berkovits and D.Sorokin, {\it Duality--symmetric
eleven--dimensional
supergravity and its coupling to M--branes} [\hepth{9711055}].

\end{thebibliography}
\end{document}